\documentclass[12pt]{article}
\usepackage{cite}
\usepackage{textcomp}
\usepackage{amsmath,amssymb,amsfonts}
\usepackage{algorithmic}
\usepackage{graphicx}
\usepackage{textcomp}
\usepackage{xcolor}
\begin{document}
{\raggedright{\huge\bf{Oscillating endoskeletal \\antibubbles}}}\\
\sloppy
Nobuki Kudo$^1$\footnote[2]{Correspondence to: Kudo@ist.hokudai.ac.jp}, Rustem Uzbekov$^{2,3}$, Ryonosuke Matsumoto$^1$, Ri-ichiro Shimizu$^1$, Craig Carlson$^4$, Nicole Anderton$^4$, Aurélie Deroubaix$^5$, Clement Penny$^5$, Albert T. Poortinga$^6$, David M. Rubin$^4$, Ayache Bouakaz$^7$, and Michiel Postema$^4$ 

($^1$Fac. Inform. Sci. \& Technol., Hokkaido Univ., Sapporo, Japan; $^2$Dept. Microscopy, Fac. Med., Univ. Tours, France; $^3$Fac. Bioeng. Inform., Moscow State Univ., Russia; $^4$School Elec. \& Inform. Eng., Univ. Witwatersrand, Johannesburg, South Africa; $^5$Dept. Int. Med., Univ. Witwatersrand, Johannesburg, South Africa; $^6$Dept. Mech. Eng., Eindhoven Univ. Technol., Eindhoven, Netherlands; $^7$Inserm U1253, Fac. Med., Univ. Tours, France)\\

{\it Submitted to the 40th Symposium on UltraSonic Electronics (USE2019).}

\section{Introduction}
An antibubble is a gas bubble with a liquid droplet inside.$^1$ Antibubbles have a potential role in ultrasonic imaging, ultrasound-guided drug delivery, and hydrocarbon leakage detection.$^1$ The liquid core droplets inside antibubbles can be stabilised with a process called Pickering stability.$^2$ Other systems proposed include spoke crystals to suspend the droplet or endoskeletal structures. In this study, the oscillatory response of microscopic endoskeletal antibubbles to ultrasound has been investigated, primarily to determine whether the presence of an endoskeleton attenuates the oscillation amplitude of the antibubble.

\section{Materials and Methods}

Three media containing (anti)bubbles were prepared for evaluation. All three media were produced as previously published.$^3$ The microscopic (anti)bubbles were stabilised using Aerosil\textsuperscript{\textregistered} R972 hydrophobised silica particles (Evonik Industries AG, Essen, Germany).$^4$ For the first medium, hereafter referred to as AB1, the aqueous cores were replaced by 2 vol\% of hydrophobically modified Zano 10 Plus zinc oxide nanoparticles 
(Umicore, Brussel, Belgium). \textbf{Fig. 1} shows an example of an AB1 endoskeletal antibubble. The silica particles have been reported to form a single elastic layer.$^4$ The zinc oxide particles forming the endoskeleton have a mean diameter of 30 nm. The inside of an AB1 endoskeletal antibubble is shown in \textbf{Fig. 2}. Two droplets, represented by darker regions, are seen to have been captured inside the endoskeleton. For the second medium, hereafter referred to as AB2, the aqueous cores were replaced by 10 vol\% of hydrophobically modified Zano 10 Plus zinc oxide. The third medium was left without cores, so it contained stabilised bubbles instead of antibubbles. This medium served as reference medium, hereafter referred to as REF. \textbf{Fig. 3} shows $z$-stacks of three (anti)bubbles. Both antibubbles are seen to contain multiple droplets, whereas the reference bubble just contains gas. For each medium, 15 mg of freeze-dried material was deposited into a FALCON\textsuperscript{\textregistered} 15 mL High-Clarity Polypropylene Conical Tube (Corning Science M\'exico S.A. de C.V., Reynosa, Tamaulipas, Mexico), after which 5 mL of 049-16797 Distilled Water (FUJIFILM Wako Pure Chemical Corporation, Chuo-Ku, Osaka, Japan) was added. 
\begin{figure}[htbp]
\centerline{
\includegraphics[width=0.75\linewidth]{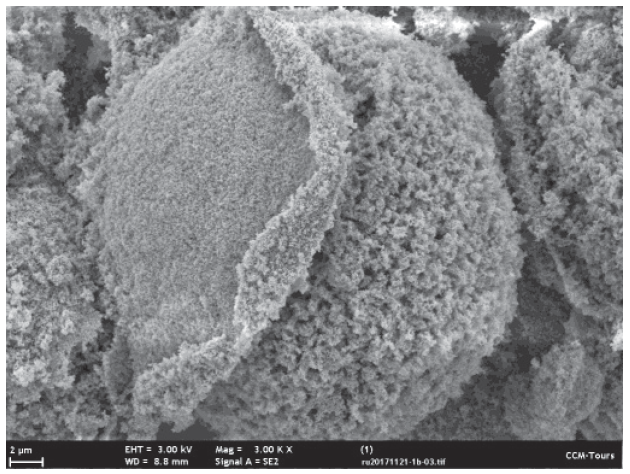}
}
\caption{Scanning electron microscope image of an AB1 antibubble. The ruptured silica membrane reveals a skeletal structure underneath.} 
\end{figure}
\begin{figure}[htbp]
\centerline{
\includegraphics[width=0.72\linewidth]{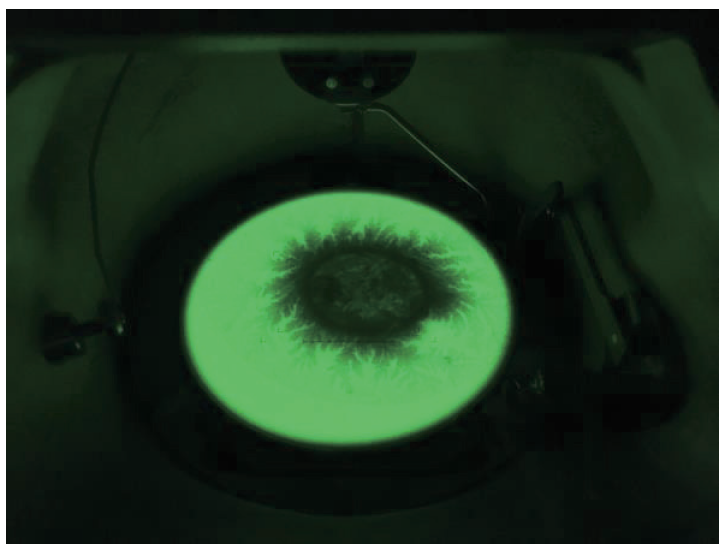}
}
\caption{Live transmission electron microscope image showing two droplets inside the AB1 endoskeleton. Salt crystals appear to have deposed on the antibubble shell.} 
\end{figure}
\begin{figure}[htbp]
\includegraphics[width=0.65\linewidth]{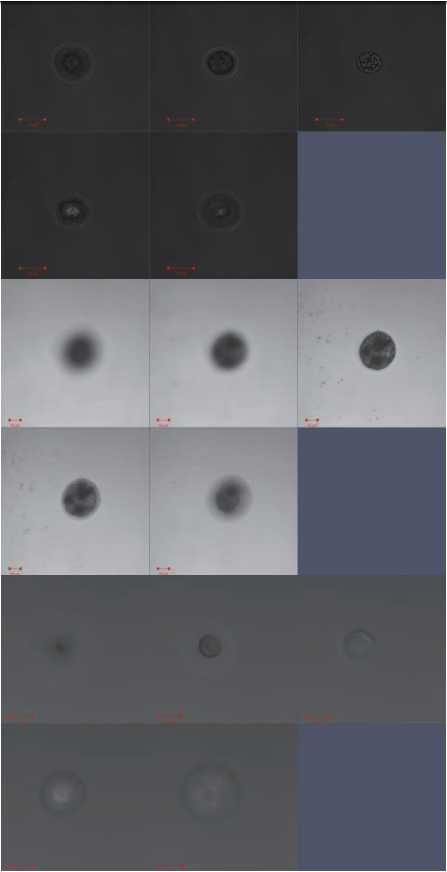}
\caption{Confocal microscopy image $z$-stacks of AB1 (top), AB2 (middle), and REF (bottom). The red bar corresponds to 10 $\mu$m.}
\end{figure}

Each emulsion was gently shaken for 1 min., after which 200 $\mu$L was pipetted into the observation chamber of a high-speed observation system.$^5$ The observation chamber was placed under an IX70 microscope (Olympus Corporation, Shinjuku-ku, Tokyo, Japan) with a LUMPlan FI/IR 40$\times$ (N.A. 0.8) objective lens. An HPV-X2 high-speed camera (Shimadzu, Nakagyo-ku, Kyoto, Japan), operating at 10 million frames per second, was couples to the microscope.$^6$ During camera recording, the materials were subjected to ultrasound pulses from a laboratory-assembled single-element transducer. Each pulse comprised 3 cycles, with a centre transmitting frequency of 1 MHz and a peak-negative pressure of 200 kPa.$^{5,6}$ The transducer?s driving signal was generated by an AFG320 arbitrary function generator (Sony-Tektronix, Shinagawa-ku, Tokyo, Japan) and
amplified by a UOD-WB-1000 wide-band power amplifier (TOKIN Corporation, Shiroishi, Miyagi, Japan). The videos recorded were segmented and analysed using MATLAB\textsuperscript{\textregistered} (The MathWorks, Inc., Natick, MA, USA).
\begin{figure}[b!!!!!!!]
\centerline{
\includegraphics[width=\linewidth]{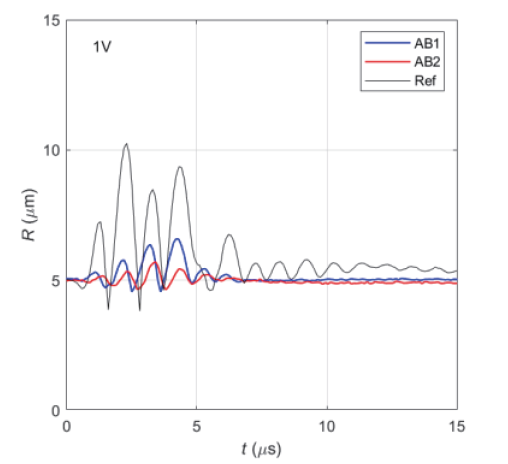}
}
\caption{Radius(time) curves of AB1, AB2, and REF.} 
\end{figure}

\section{Results and Discussion}
\textbf{Fig. 4} shows three radius(time) curves of representative (anti)bubbles with equilibrium radii of 5 $\mu$m, selected from 10 high-speed video recordings, each consisting of 256 frames. The bubble, REF, represented by the black line, is seen to oscillate with the highest amplitudes, whereas the antibubble with the highest volumetric endoskeleton content, AB2, represented by the red line, has the lowest oscillation amplitudes. Similar results were observed in the other video recordings (not shown). It is also noted that the antibubble oscillations fade within 3 $\mu$s after sonication stops, whereas the reference bubble oscillates for more than 10 $\mu$s after sonication stops. This indicates that the endoskeleton has damping properties. The possibility of acoustic energy storage in the endoskeleton.

\section{Conclusions}
The presence of an endoskeleton with droplets hampers the oscillation of antibubbles as compared to identical bubbles without endoskeleton and droplet content. 

\section*{References}
\begin{enumerate}
\item	Y. Vitry, S. Dorbolo, J. Vermant and B. Scheid: Adv. Colloid Interface Sci. 270 (2019) 73.
\item	A.T. Poortinga: Langmuir 27 (2011) 2138.
\item	A.T. Poortinga: Colloid. Surf. A 419 (2013) 15.
\item	M. Postema, A. Novell, C. Sennoga, et al.: Appl. Acoust. 137 (2018) 148.
\item	N. Kudo: IEEE Trans. Ultrason. Ferroelect. Freq. Control 64 (2017) 273.
\item	S. Imai, N. Kudo: Proc. IEEE Ultrason. Symp. (2018) 8579713.
\end{enumerate}
\end{document}